# Observation of Ultrahigh Mobility Surface States in a Topological Crystalline Insulator by Infrared Spectroscopy

Ying Wang[1], Guoyu Luo[2], Junwei Liu[3], R. Sankar[4], Nan-Lin Wang[5], Fangcheng Chou[4], Liang Fu[3], Zhiqiang Li[2*]

[1]National High Magnetic Field Laboratory, Tallahassee, Florida 32310, USA.

[2]College of Physical Science and Technology, Sichuan University, Chengdu, Sichuan 610064, China.

[3]Department of Physics, Massachusetts Institute of Technology, Cambridge, Massachusetts 02139, USA.

[4]Center for Condensed Matter Sciences, National Taiwan University, Taipei 10617, Taiwan.

[5]International Center for Quantum Materials, School of Physics, Peking University, Beijing 100871, China.

*e-mail: zhiqiangli@scu.edu.cn

**Topological crystalline insulators (TCIs) possess metallic surface states protected by crystalline symmetry, which are a versatile platform for exploring topological phenomena and potential applications. However, progress in this field has been hindered by the challenge to probe optical and transport properties of the surface states owing to the presence of bulk carriers. Here we report infrared (IR) reflectance measurements of a TCI, (001) oriented $Pb_{1-x}Sn_xSe$ in zero and high magnetic fields. We demonstrate that the far-IR conductivity is unexpectedly dominated by the surface states as a result of their unique band structure and the consequent small IR penetration depth. Moreover, our experiments yield a surface mobility of 40,000 $cm^2 V^{-1} s^{-1}$, which is one of the highest reported values in topological materials, suggesting the viability of surface-dominated conduction in thin TCI crystals. These findings pave the way for exploring many exotic transport and optical phenomena and applications predicted for TCIs.**

Recently, a new class of insulators called topological crystalline insulators (TCIs) were predicted and observed in IV−VI semiconductors[1-8], which have attracted tremendous scientific interest[8-17]. These materials host gapless Dirac-like surface states (SS) that are protected by crystalline symmetry[1,2] instead of time-reversal symmetry[18-20]. Consequently, breaking the crystal symmetry can open a bandgap in these SS[3, 12], offering new opportunities for bandgap engineering by strain or structural distortion. Moreover, compared with $Z_2$ topological insulators[18-20], the (001) SS of TCIs have been predicted to exhibit a wider range of tunable electronic properties under many types of perturbations breaking the crystalline symmetry, such as interface superconductivity[9], spin-filtered edge states[10], quantum anomalous Hall effect[11], Weyl fermions[13] and valley-dependent optical properties[14]. Therefore, TCIs are emerging as a very versatile material system not only for exploring topological quantum phenomena, but also for potential device applications in electronics, spintronics and optoelectronics[8-14]. Several novel characteristics of TCIs have been revealed by surface sensitive probes[3-6, 12]. However, the transport and optical properties of the (001) SS, which are fundamentally important and arguably most relevant to applications[8-14], have remained especially challenging to measure because of the overwhelming effects of bulk carriers in previous studies[15-17]. This has seriously hampered the progress in this field.

Here we present IR reflectance measurements of a TCI, $Pb_{1-x}Sn_xSe$ (x=0.23-0.25) single crystals with (001) surface in zero and high magnetic fields. From the bandgap, Fermi velocity and Fermi energy of the bulk bands determined from our data, the bulk Drude spectral weight can be estimated, which is found to be much less than the measured Drude weight, indicating substantial contributions from surface carriers. Secondly, the spectral features in magneto-reflectance spectra below 25 meV can be attributed to a dominant resonance at $\omega_c^{ss} \propto B$ based on theoretical study of cyclotron resonance (CR) of the SS, the frequency of which obtained from our data ($\omega_c^{ss}$) is quantitatively consistent with those estimated from previous scanning tunneling microscopy (STM) and angle-resolved photoemission spectroscopy (ARPES) experiments. Moreover, the spectral weight of the dominant resonance in magnetic field is in accord with the extra Drude weight in zero field besides the bulk contribution. Above all, we demonstrate that the resonance at $\omega_c^{ss}$ in field is well below the energy range of all Landau level (LL) transitions (including CR) from the bulk states, so it can only be assigned to the SS. Therefore, all these findings taken together provide robust evidence for SS in $Pb_{1-x}Sn_xSe$.

Remarkably, we find that the IR conductivity of $Pb_{1-x}Sn_xSe$ is dominated by the SS in the far-IR range (7-25 meV or 2-6 THz) despite the presence of bulk carriers. We show that this unexpected property arises from the unique band structure of the SS (Fig. 1a) and the resultant high surface carrier density and small IR penetration depth. Furthermore, our experiments yield a surface mobility of ~40,000 $cm^2 V^{-1} s^{-1}$ based on analysis of CR mode of the SS, which is 1-2 orders of magnitude higher than that in TCI thin films[21-23] and among the highest reported values in topological materials. The ultrahigh surface mobility and other transport parameters obtained here suggest that surface-dominated transport can be achieved in (001) oriented $Pb_{1-x}Sn_xSe$ crystals with sub-micron thickness. Our findings open up new opportunities for exploring many exotic transport and optical phenomena and applications predicted for TCIs[8-14], ranging from quantum anomalous Hall effect and Weyl fermions to spintronics and valleytronics.

**Results**

In this work, $Pb_{1-x}Sn_xSe$ single crystals with an actual composition of x=0.23-0.25 were investigated, which are in the TCI phase and host gapless SS[4] (Fig. 1a). The actual composition is determined from the bulk bandgap[4] as discussed below. These materials have a direct bulk bandgap located at four L points (valleys) in the three-dimensional Brillouin zone. Our samples are n-doped with the Fermi energy $E_F$ in the bulk conduction band[3]. The IR reflectance spectra $R(\omega)$ of $Pb_{1-x}Sn_xSe$ crystals with (001) surface were measured in zero field and in magnetic field applied perpendicular to the surface of the samples. Figure 1 depicts the $R(\omega)$ spectrum and the dissipative part of the optical conductivity $\sigma_1(\omega)$ in zero field at T=8K (see Methods). The $R(\omega)$ spectrum shows a typical metallic behavior: $R(\omega)$ is very close to 1 below 15 meV; at higher energy the reflectance is gradually depressed towards a plasma minimum at about 30 meV, followed by a peak around 120 meV. The energy of the plasma minimum observed here is much lower than those in earlier IR studies[16,17], indicating much lower total carrier density in our samples due to low density of Se vacancy defects (see Methods and Supplementary Discussion 1). The zero field $\sigma_1(\omega)$ spectrum exhibits a Drude component below 25 meV and a threshold-like feature around 100 meV. The narrow plasma minimum in $R(\omega)$ with a half width ~3 meV is a direct manifestation of the very narrow Drude peak in $\sigma_1(\omega)$, which is corresponding to a very low scattering rate. The threshold feature in $\sigma_1(\omega)$ can be assigned to the onset of interband transitions for the bulk around $E_{inter} \approx$ $\Delta + E_F(1 + \frac{m_c}{m_v})$ as illustrated in the inset of Fig. 1c, where $\Delta$ is the bulk bandgap, the Fermi energy $E_F$ is defined with respect to the bottom of the conduction band, $m_c$ and $m_v$ are effective masses of the conduction and valence band, respectively. The experimental absorption coefficient spectrum (not shown) also exhibits a similar threshold feature in the same energy range. It is shown that in doped semiconductors the absorption coefficient can be written as[24,25] $\alpha(\omega) \propto \alpha_0(\omega)[1 - f(\omega, E_F, T)]$, where $\alpha_0(\omega) \propto \sqrt{\hbar\omega - \Delta}/\hbar\omega$ is the absorption coefficient for the undoped material and $f(\omega, E_F, T) = [1 + \exp(\frac{\hbar\omega - E_{inter}}{(1+\frac{m_c}{m_v}) k_B T})]^{-1}$ is the Fermi distribution. Fitting the experimental absorption coefficient using the equation above, we find $E_{inter} \sim 100$ meV as indicated by the dashed line in Fig. 1c.

More insights into the band structure of Pb$_{1-x}$Sn$_x$Se can be provided by a systematic investigation of the magneto-reflectance $R(\omega, B)$ spectra displayed in Fig. 2 (see Methods). Strikingly, the zero field $R(\omega)$ spectrum is strongly modified by a magnetic field. The $R(\omega, B)$ spectra are strongly suppressed below 25 meV in magnetic fields and no longer extrapolate to unity in the limit of $\omega \to 0$. Moreover, a series of new resonance features are observed in $R(\omega, B)$ extending up to 300 meV, all of which evolve systematically with magnetic fields. In order to acquire a complete understanding of the features in $R(\omega, B)$ spectra, the optical conductivity in magnetic field was extracted from an analysis of $R(\omega, B)$ using the magneto-Drude-Lorentz model[26-28] (see Methods). Fig. 3a depicts the real part of the optical conductivity $Re\ \sigma_{xx}(\omega, B)$ from 50 meV to 300 meV, which exhibit several resonance (absorption) peaks systematically shifting to higher energies with increasing magnetic field. The energies ($E$) of the absorption features in $Re\ \sigma_{xx}(\omega, B)$ are displayed in Fig. 3b, which show good agreement with the energies of the resonance features in $R(\omega, B)$. The field-dependent absorption features can be assigned to LL transitions. Interestingly, the observed transitions are not equally spaced in energy in any spectrum, which is in stark contrast to the behavior of systems with quadratic energy-momentum dispersion. We find that the observed resonance features in $Re\ \sigma_{xx}(\omega, B)$ can be well described by LL transitions of three-dimensional (3D) massive Dirac fermions[8, 15], whose LLs have the form:

$$E_n(k_z) = \pm\delta_{n,0}\sqrt{(\Delta/2)^2 + (\hbar v_F k_z)^2} + sgn(n)\sqrt{2e\hbar v_F^2 B|n| + (\Delta/2)^2 + (\hbar v_F k_z)^2} \quad (1)$$

which have two zeroth LLs labeled as $n=+0$ and $n=-0$. Here, the integer $n$ is LL index, $k_z$ is momentum along the direction of the magnetic field, $e$ is the elementary charge, $\hbar$ is Planck's constant divided by $2\pi$, $v_F$ is the Fermi velocity, $\delta$ is the Kronecker delta function, and $sgn(n)$ is the sign function. Theoretical studies showed that the optical conductivity has sharp peaks at energies of LL transitions at $k_z = 0$ because of singularities in the joint-density-of-states between LLs at these energies[29]. From the selection rule[29, 30] for allowed optical transitions from $LL_n$ to $LL_{n'}$, $\Delta n = |n| - |n'| = \pm 1$, we find that all the observed resonances can be assigned to allowed LL transitions based on equation (1) with $k_z = 0$, $\Delta \approx 64 \pm 3$ meV and $v_F \approx (0.400 \pm 0.005) \times 10^6$ m s$^{-1}$ (Fig. 3b and 3c), which are determined from least squares fit of the observed transition energies. This analysis shows that the sharp resonances in Fig. 3a mainly arise from massive Dirac fermions of the bulk states[8, 15]. Since the Sn substitution level directly leads to changes in the bulk bandgap, the actual composition of our samples can be determined from $\Delta$ to be x=0.23-0.25 (ref. 4).

Our data allow us to identify the signatures of SS in Pb$_{1-x}$Sn$_x$Se because all spectroscopic features of the bulk states can be determined. The first evidence for the SS is from an analysis of the Drude spectral weight (area under $\sigma_1(\omega)$) in zero field. The low energy bulk states can be described by massive Dirac fermions: $E(k) = \pm\sqrt{\hbar^2 v_F^2 k^2 + (\Delta/2)^2}$, with $\Delta$ and $v_F$ given above. We can estimate $E_F$ of the bulk states from $E_{inter} \approx \Delta + 2E_F$ (note that $m_c = m_v$ for massive Dirac fermions) shown in Fig. 1c. Alternatively, the transition $LL_{-0} \to LL_{+1}$ disappears at $B<6T$ (Fig. 2 and 3), so the Fermi energy can be estimated as $E_F \sim LL_{+1}(B=6T) - \Delta/2$ using equation (1). Either method yields $E_F \approx 17 \pm 2$ meV. From the band dispersion and $E_F$, we obtain $k_F \approx 0.014 \pm 0.001$ Å$^{-1}$. Recent transport studies found that the bulk Fermi surface pockets of Pb$_{1-x}$Sn$_x$Se in the TCI phase are nominally spherical[15], so the bulk carrier density can be estimated by $n_{bulk} = \frac{1}{(2\pi)^3}\frac{4}{3}\pi k_F^3 g_s g_v \approx (3.7 \pm 1.1) \times 10^{17}$ cm$^{-3}$, where $g_s = 2$ and $g_v = 4$ are the bulk spin and valley degeneracy, respectively. Similarly low carrier density has been reported by previous transport experiments[15]. Moreover, the effective mass of massive Dirac fermions at $E_F$ can be calculated from $m_{bulk} = \frac{E_F + \Delta/2}{v_F^2} \approx 0.054 \pm 0.003$ $m_e$, where $m_e$ is bare electron mass. The spectral weight (SW) of the Drude absorption is related to the bare plasma frequency $\omega_P^2 = 4\pi e^2 n/m$ by[31]:

$$SW = \int_0^{\omega_0} \sigma_1(\omega)d\omega = \frac{\pi \Omega^{-1}}{120}\omega_P^2 \quad (2)$$

where $\omega_0$ is a cut-off frequency separating the Drude component from the interband transitions, $\Omega$ is Ohm, $\sigma_1$ is in $\Omega^{-1}$cm$^{-1}$ and all frequencies are in cm$^{-1}$. From $n_{bulk}$, $m_{bulk}$ and equation (2), we estimate that $SW_{bulk} \approx (1.6 \pm 0.5) \times 10^4$ $\Omega^{-1}$cm$^{-2}$ for the bulk states from calculating $\omega_{P,bulk}$. On the other hand, the total Drude spectral weight can be determined by the total plasma frequency $\omega_P$ of the Drude component in the experimental data using $SW_{total} = \frac{\pi \Omega^{-1}}{120}\omega_P^2$, where $\omega_P$ is related to the screened plasma frequency $\widetilde{\omega}_P$ by $\omega_P = \widetilde{\omega}_P\sqrt{\varepsilon_\infty}$, with $\widetilde{\omega}_P \sim 260$ cm$^{-1}$ (32.4 meV) corresponding to the plasma minimum in $R(\omega)$ and $\varepsilon_\infty$ representing all electronic contributions to the dielectric constant other than the Drude component. Therefore, the total plasma frequency $\omega_P$ can be obtained from $\varepsilon_\infty$, which is determined by all Lorentzian oscillators obtained from Drude-Lorentz fit of the *entire* R(ω) spectrum (see Supplementary Discussion 5). We find that $\varepsilon_\infty \sim 45 \pm 9$, so $SW_{total} \approx (7.9 \pm 1.6) \times 10^4$ $\Omega^{-1}$cm$^{-2}$. A comparison between $SW_{bulk}$ and $SW_{total}$ shows that the bulk contribution can only account for about 20% of the total Drude spectral weight of the system. We emphasize that the total Drude spectral weight and that from bulk states are both accurately determined from our data, which strongly suggests dominant surface contribution to the total Drude absorption in zero field.

The low energy magneto-reflectance of Pb$_{1-x}$Sn$_x$Se provides further evidence for the SS. The systematic suppression of $R(\omega, B)$ spectrum with $B$ field below 25 meV suggests that the system becomes more insulating with increasing field. This is reminiscent of the typical behavior of CR. For instance, similar low energy behaviors in $R(\omega, B)$ were also observed in graphite[27] due to CR, which manifests itself in $Re\ \sigma_{xx}(\omega)$ as a peak at energies of $E \propto B$. We now show that the spectral feature below 25 meV in $R(\omega, B)$ and its evolution with $B$ field are unambiguous signatures of CR from the SS (Supplementary Discussion 2 and 3). As shown in Fig. 1a and 4a, the band structure of SS on the (001) surface features two generations of Dirac fermions[7], starting from a pair of Dirac cones located at the $\bar{X}$ points of the surface Brillouin zone with their Dirac points at $E_{H1}^{DP}$ and $E_{H2}^{DP}$. The hybridization between these two Dirac cones leads to a gap in all directions except along the $\overline{\Gamma X}$ line (Fig. 1a), where a pair of Dirac points exist that are protected by the (110) mirror symmetry[2,7]. Recent experiments and theoretical calculations[3,32,33] suggest that $E_{H1}^{DP}$ is close to $E_F$ of the bulk states for Pb$_{1-x}$Sn$_x$Se. In a magnetic field perpendicular to the surface, it is shown[32] that the LLs of the SS near $E_F$ are well approximated by LLs of two independent Dirac cones at $E_{H1}^{DP}$ and $E_{H2}^{DP}$ (Fig. 4b), which have energies of $E_n = sgn(n)\sqrt{2e\hbar(\bar{v}_F^{SS})^2 B|n|}$ with respect to each Dirac point and $\bar{v}_F^{SS} \sim 0.4 \times 10^6$ m s$^{-1}$ (ref. 3, 32,33). Within this picture, a CR mode due to the intraband LL transition associated with the Dirac cone at $E_{H2}^{DP}$ is expected at low energy (Fig. 4b). The surface Fermi energy with respect to $E_{H2}^{DP}$ is estimated to be $E_F^{SS} \sim 136 \pm 14$ meV based on STM and ARPES experiments[3, 32,33], from which the energy of the surface CR mode is expected to be $\omega_c^{SS} = eB/m_{ss}$ with $m_{ss} = E_F^{SS}/(\bar{v}_F^{SS})^2 = 0.15 \pm 0.015\ m_e$. LL transitions of SS associated with the Dirac cone at $E_{H1}^{DP}$ can also contribute to the *R(ω, B)* spectra at $\sqrt{2e\hbar(\bar{v}_F^{SS})^2 B}$ or higher energy, but it is very difficult to separate them from bulk LL transitions that are in the same energy range. It is shown[32] that LLs associated with the Dirac cone along the $\overline{\Gamma X}$ line are restricted in the energy range between $E_{VHS1}$ and $E_{VHS2}$ in Fig. 4a and 4b, so these LLs below $E_F$ cannot be probed from optical measurements, which requires optical transitions from a LL below $E_F$ to one above $E_F$. To examine the IR signatures of the surface CR mode, the magneto-Drude-Lorentz model is used to simulate the *R(ω, B)* spectra (see Methods). The real and imaginary parts of $\sigma_{xx}(\omega)$, $\sigma_{xy}(\omega)$, $\sigma_+(\omega)$ and $\sigma_-(\omega)$ in representative models are displayed and discussed in Supplementary Discussion 2. We find that the principle features of *R(ω, B)* can be reproduced by model optical conductivity spectra that are based on the expected behaviors of the surface CR mode[32]: remarkably, a strong resonance peak at $\omega_c^{ss}$ in $Re\ \sigma_{xx}(\omega, B)$ that changes linearly with *B* field (figure 4c) can capture the spectral feature in *R(ω, B)* below 25 meV and its evolution with *B* field. In particular, the analysis on our data yields $m_{ss}$ values in the range of 0.15-0.19 $m_e$ for the CR mode, which are consistent with those estimated from STM and ARPES experiments[3,33] within 15% as shown in figure 4d (see Supplementary Discussion 2). The small deviation between our results and STM and ARPES measurements may arise from the difference in Fermi energy in different samples. Based on LL energies calculated from $\Delta$ and $v_F$ for the bulk and equation (1), the peak between 30 and 50 meV in $Re\ \sigma_{xx}(\omega)$ spectra can be assigned to the $LL_{+0} \to LL_{+1}$ transition (intra-band LL transition or CR) from the bulk states for B>6T, which is responsible for the feature around 37 meV in *R(ω, B)*. Because $E_F$ is between $LL_{+0}$ and $LL_{+1}$ above B~6T, the bulk CR $LL_{+0} \to LL_{+1}$ around 37 meV has the lowest energy among all allowed LL transitions for the bulk for B>6T (Fig 3c). Therefore, the spectral feature in *R(ω, B)* below 25 meV can only be assigned to the SS, because it is well below the energy range of *all* LL transitions (including CR) from the bulk states for B>6T. Moreover, from the area under the resonance at $\omega_c^{ss}$ in $Re\ \sigma_{xx}(\omega)$, the spectral weight of this resonance is found to be $SW_{SS} \sim (4.4 \pm 0.9) \times 10^4 \ \Omega^{-1} \text{cm}^{-2}$, which is in agreement with the extra Drude spectral weight in zero field besides the bulk contribution $\sim (6.3 \pm 2.1) \times 10^4 \ \Omega^{-1} \text{cm}^{-2}$. This agreement provides further support for our identification of the surface CR mode. Therefore, our data in zero and high magnetic fields taken together provide robust evidence for the IR signatures of SS in Pb$_{1-x}$Sn$_x$Se.

Strikingly, the far-IR conductivity of (001) oriented Pb$_{1-x}$Sn$_x$Se is dominated by the SS in the energy range of the Drude absorption in zero field and that of the surface CR mode in magnetic field, which is around 7-25 meV or 2-6 THz. We now show that this is a consequence of the unique band structure of the (001) SS. The SS of our n-doped samples has two Fermi surfaces associated with the two Dirac cones at $E_{H1}^{DP}$ and $E_{H2}^{DP}$ (Fig. 4a). In particular, the latter Dirac cone gives rise to a large Fermi surface with $\bar{k}_F^{SS} \sim E_F^{SS}/(\hbar \bar{v}_F^{SS}) \sim 0.052 \pm 0.005 \ \text{Å}^{-1}$ (ref. 3,32,33). Neglecting the small contribution from the Fermi surface associated with the Dirac cones at $E_{H1}^{DP}$ and taking into account $g_s = 1$ and $g_v = 2$ for the SS, we estimate that the large SS Fermi surface leads to a very high surface carrier density $n_{SS}^{2D} \sim (4.2 \pm 0.8) \times 10^{12} \ \text{cm}^{-2}$, which is more than one order of magnitude higher than that of the (111) surface in TCIs[21, 34]. The large value of $n_{SS}^{2D}$ of the (001) surface is critical for achieving surface dominated far-IR conductivity in Pb$_{1-x}$Sn$_x$Se despite the presence of bulk carriers. Our measurements only probe the region of the sample within the IR penetration depth $\delta$. An estimate for the length scale of $\delta$ for our samples between 7-25 meV can be obtained from a spectral weight analysis. The Drude weight of the (001) SS can be estimated as $e^2 E_F^{SS} g_s g_v / (8\hbar^2)$ (ref. 35), which only includes contributions from the massless Dirac fermions associated with the Dirac cone at $E_{H2}^{DP}$. The same surface Drude weight can be obtained from $SW_{SS}\delta$ based on our IR measurements, so the *averaged* length scale of $\delta$ between 7-25 meV is estimated to be $13 \pm 4 \ nm$ for our samples. It can be shown that $\delta$ is comparable to the penetration depth of the SS (see Methods). This suggests that the high carrier density of the (001)

SS leads to strong IR absorption and attenuates the IR light within a very small penetration depth, which effectively reduces the bulk contribution to the far-IR conductivity. It is instructive to compare our results with a recent IR study of (111)-oriented TCI thin films[34], which reported substantially lower surface spectral weight compared to the bulk contribution. It is found that the (111) surface carrier density is ten times lower and the IR penetration depth is much larger than those in our samples[34], owing to a much smaller $k_F$ for the (111) SS. These results are consistent with our findings that the high surface carrier density reduces the IR penetration depth and thus the bulk contribution to the overall far-IR conductivity in (001)-oriented $Pb_{1-x}Sn_xSe$. Therefore, the unique band structure of (001) SS plays a crucial role in achieving surface dominated far-IR conductivity observed in our study. The large difference in the IR properties of (001) and (111) surfaces is a new aspect in TCIs compared to those in $Z_2$ topological insulators[36-39].

Our study sheds new light on the mobility of the (001) SS of $Pb_{1-x}Sn_xSe$. From half width of the surface CR mode in $Re\ \sigma_{xx}(\omega)$, the scattering rate of the SS is estimated to be $1/\tau_{SS} \sim 1.2 \pm 0.6$ meV, which clearly manifest itself as the very narrow dip feature around 32 meV in $R(\omega, B)$ (Supplementary Discussion 4). Therefore, the mobility of the SS can be estimated: $\mu_{SS} = e\bar{v}_F^{SS}\tau_{SS}/(\hbar\bar{k}_F^{SS}) = e\tau_{SS}/m_{ss} \sim 40,000$ cm$^2$ V$^{-1}$ s$^{-1}$ with an uncertainty of about 50%. Such a high surface mobility is very promising for studying the intrinsic physics of TCIs. Detailed analysis on the scattering rate and mobility from previous IR study of LL transitions in graphene[40] support our estimation of surface mobility in TCIs since both systems feature massless Dirac fermions (Supplementary Discussion 4). In contrast, previous studies on TCI thin films reported surface mobility values in the range of 100-2,000 cm$^2$ V$^{-1}$ s$^{-1}$ (ref. 21-23) and a broad surface IR absorption feature that is ten times wider than that in our data[34]. The low mobility values of these TCI thin films are very likely limited by sample quality. We note that the small scattering rate and high mobility of the (001) SS also play an important role in leading to strong surface absorption in a narrow energy range in far-IR and thus a small IR penetration depth, which is responsible for the observed surface-dominated IR conductivity.

A long-standing challenge in the research of TCIs as well as $Z_2$ topological insulators is to realize surface-dominated transport[41-45]. Our study elucidates the surface and bulk contributions to electronic transport. From the effective mass of bulk states $m_{bulk}$ obtained from zero field data and the scattering rate estimated from bulk LL transitions $1/\tau_{bulk} \sim 10$ meV, the bulk mobility of our samples can be estimated: $\mu_{bulk} \sim 13,000$ cm$^2$ V$^{-1}$ s$^{-1}$. Similarly high mobility has been reported by previous transport experiments[15]. For a sample with thickness $d$, the fraction of surface contribution to the total conductance can be estimated as $n_{SS}^{2D}\mu_{SS}e/(n_{SS}^{2D}\mu_{SS}e + n_{bulk}d\mu_{bulk}e)$. The mobility and carrier density values obtained in our study suggest that the SS in our samples will contribute to more than 50% of the total conductance in crystals with $d$<0.4 μm. Surface-dominated transport can be achieved in even thicker crystals if the bulk carrier density is reduced further in future.

In conclusion, our study has revealed several IR signatures of SS in Pb$_{1-x}$Sn$_x$Se including their contribution to the Drude absorption in zero field and their CR mode in magnetic field, whose spectral weights are further found to be consistent. Moreover, the frequency of the surface CR mode $\omega_c^{SS}$ inferred from our data is in accord with those estimated from theoretical studies and STM and ARPES measurements. Finally, we show that the resonance at $\omega_c^{SS}$ in field can only be assigned to the SS, because it is well below the energy range of all bulk LL transitions. Our study demonstrates the viability of surface-dominated IR conductivity and transport as well as an ultrahigh surface mobility of ~40,000 cm$^2$ V$^{-1}$ s$^{-1}$ in Pb$_{1-x}$Sn$_x$Se with (001) surface. We expect that these results can be found in a broad class of IV–VI semiconductor TCIs Pb$_{1-x}$Sn$_x$Se(Te) because of the similar properties of their SS[1-8]. These findings establish Pb$_{1-x}$Sn$_x$Se(Te) thin crystals (potentially with electrostatic gating) as a fertile system for accessing the optical and transport properties of the (001) SS, which is a crucial step for exploring many novel topological phenomena in TCIs and their potential electronic, spintronic and valleytronic applications[8-14].

**Methods**

**IR measurements and data analysis**

$Pb_{1-x}Sn_xSe$ (x=0.23-0.25) single crystals with centimeter size were grown by the Bridgman growth method. As $Pb_{1-x}Sn_xSe$ crystals have cubic symmetry, the samples in our study have been chosen from naturally cleaved cubic crystals with easy low indexing (00L) cleavage planes. The carrier density in the crystals is batch (growth) dependent even for the same nominal Pb/Sn ratio because of the self-doping effect of Se vacancy defects (see Supplementary Discussion 1). The crystals used in our IR study are from a batch with low carrier (defect) density. The samples were cleaved in ambient conditions before the first IR measurement, and their exposure time in air was minimized between subsequent measurements. The zero field reflectance spectrum $R(\omega)$ was measured in the energy range of 7.5–900 meV at a temperature of T=8K. The magneto-reflectance ratio spectra $R(\omega, B)/R(\omega, B=0)$ were measured at T~4.5K in a superconducting magnet in the Faraday geometry (magnetic field perpendicular to the sample surface). The $R(\omega, B)$ spectra were obtained by multiplying the magneto-reflectance ratio spectra by $R(\omega)$ at zero field (Supplementary Discussion 1). We estimate the experimental uncertainty of the absolute values of $R(\omega, B)$ is about 1%-1.5%, but the relative change (evolution) of $R(\omega, B)$ with magnetic field has a smaller uncertainty owing to the better accuracy of the magneto-reflectance ratio data.

As an approximation for exploring spatially averaged properties of the SS, our data analysis employs a model with uniform carrier density $n_{SS}$ for the SS without a sharp surface/bulk interface, which will be elaborated below. In such a model, the IR data can be analyzed using formulas for bulk materials and the resulting optical conductivity contains contributions from both the SS and bulk, with all quantities of the SS being spatially averaged ones along the direction perpendicular to the surface within the IR penetration depth. The complex optical conductivity $\sigma(\omega) = \sigma_1(\omega) + i\sigma_2(\omega)$ in zero field at 8K was obtained from analysis of $R(\omega)$ using Drude-Lorentz model combined with Kramers-Kronig (KK) constrained variational dielectric functions[28]. Alternatively, $\sigma(\omega)$ in zero field can be obtained from KK transformation of $R(\omega)$ data, in which Hagen-Rubens formula was employed in the low energy region and the $R(\omega)$ data at 8K was extend to 5.765 eV using room temperature reflectance data. The two approaches above yield consistent $\sigma(\omega)$ spectra with identical spectral features. In magnetic field, the $R(\omega, B)$ spectra were analyzed using magneto-Drude-Lorentz model[26], in which the optical conductivity for right- and left- circularly polarized light (denoted "+" and "-" respectively) are given by:

$$\sigma_\pm(\omega) = \frac{\omega}{4\pi i}\left(\varepsilon_\infty^B - 1 + \sum \frac{\omega_{p,n}^2}{\omega_{0,n}^2 - \omega^2 - i\gamma_n\omega \mp \omega_{c,n}\omega}\right) \quad (3)$$

where $\omega_{p,n}$, $\omega_{o,n}$, $\gamma_n$ and $\omega_{c,n}$ are the plasma frequency, energy, linewidth and cyclotron energy of the n-th oscillator, respectively. $\varepsilon_\infty^B$ represents high-energy contributions to the dielectric constant from energy range higher than all the oscillators. In our analysis, we choose $\omega_{o,n} = 0$ for all oscillators. The magneto-Drude-Lorentz model ensures that the real and imaginary parts of optical conductivity are constrained by Kramers–Kronig relations, therefore it is commonly used to describe LL transitions including cyclotron resonance. The xx and xy components of the optical conductivity are: $\sigma_{xx}(\omega) = [\sigma_+(\omega) + \sigma_-(\omega)]/2$, $\sigma_{xy}(\omega) = [\sigma_+(\omega) - \sigma_-(\omega)]/2i$. The dielectric function is $\varepsilon_\pm(\omega) = 1 + \frac{4\pi i \sigma_\pm(\omega)}{\omega}$. The complex reflectivity is given by $r_\pm(\omega) = \frac{1-\sqrt{\varepsilon_\pm(\omega)}}{1+\sqrt{\varepsilon_\pm(\omega)}}$. Finally, the measured reflectance spectrum is given by:

$$R(\omega) = \frac{|r_+(\omega)|^2 + |r_-(\omega)|^2}{2} \quad (4)$$

The experimental $R(\omega,B)$ data were fit with model reflectance spectrum calculated from Eq.(4). In principle, a complete determination of both $\sigma_+(\omega)$ and $\sigma_-(\omega)$ cannot be achieved without Kerr rotation measurements. However, after trying numerous versions of $\sigma_+(\omega)$ and $\sigma_-(\omega)$ to fit $R(\omega,B)$, we find that good fits always yield quite similar $Re\ \sigma_{xx}(\omega)$ spectra in the energy range of 75-350 meV. Qualitatively, because of the small changes in *R(ω, B)/R(ω, B=0)* (less than 5%) and the relatively flat spectral feature in *R(ω, B=0)* in 75-350 meV, the resonances in *R(ω, B)* are mathematically related to peaks in $Re\ \sigma_{xx}(\omega)$ at very similar energies, which allows us to determine $Re\ \sigma_{xx}(\omega)$ with good accuracy in this energy range. On the other hand, due to the strong energy dependence of the spectral features in *R(ω, B=0)* below 50 meV and much larger changes in *R(ω, B)/R(ω, B=0)* (~20%), it is impossible to accurately determine $Re\ \sigma_{xx}(\omega)$ in the low energy range. Therefore, we use one oscillator at $\omega_c^{SS}$ based on theoretical results on the CR of the SS in TCIs[32] and several much weaker oscillators to simulate the *R(ω, B)* spectra below 60 meV, employing the magneto-Drude-Lorentz model. While our discussions are mainly focused on the $Re\ \sigma_{xx}(\omega)$ spectrum, the real and imaginary parts of $\sigma_{xx}(\omega)$, $\sigma_{xy}(\omega)$, $\sigma_+(\omega)$ and $\sigma_-(\omega)$ in representative models are displayed and discussed in Supplementary Discussion 2.

Our model is a good approximation for investigating $Pb_{1-x}Sn_xSe$ (x=0.23-0.25), which has a large SS penetration depth. The length scale of the penetration depth (spatial spread) of the SS into the bulk can be estimated as $\lambda_{SS} \sim \hbar v_F^{SS}/\Delta$, where $v_F^{SS}$ is the Fermi velocity of the Dirac fermions of the SS and $\Delta$ is the bulk bandgap[12]. $\lambda_{SS}$ can be large for small values of $\Delta$ and even goes to infinity with $\Delta = 0$. The value of $\lambda_{SS}$ in $Pb_{1-x}Sn_xSe$ can be estimated from a comparison with topological insulator $Bi_2Se_3$, whose $\lambda_{SS}$ has been found to be ~2.5 nm (2.5 quintuple layers) (ref. 46). The bulk bandgap for our sample is about 1/5 of that of $Bi_2Se_3$ and the two materials have quite similar $v_F^{SS}$ values[46], so the length scale of $\lambda_{SS}$ is expected to be ~12.5 nm. Moreover, theoretical calculations[12] have shown that the percentage of the SS wavefunction in the topmost layer for Pb$_{1-x}$Sn$_x$Se (x=0.23-0.25) is ~0.5%, which suggests that $\lambda_{SS}$ is much larger than 200 layers or 60 nm based on the lattice constant ~0.3nm. On the other hand, the far-IR range of 7-25 meV is the main focus of our discussions on the SS, where the *averaged* length scale of the IR penetration depth estimated from our data is $\delta \sim 13 \pm 4$ nm, which is comparable to (or smaller than) the length scale of $\lambda_{SS}$. Our analysis employs a model with two main approximations: (1) Both the SS and bulk states exist within the entire IR penetration depth without a sharp surface/bulk interface, which is justified since $\delta \sim \lambda_{SS}$ in Pb$_{1-x}$Sn$_x$Se in the frequency range of interest; (2) As an approximation for studying spatially averaged properties of the SS, the model assumes a uniform $n_{SS}$ for the SS. With these approximations the IR data can be analyzed using the methods detailed above. Regarding SS penetration depth, Pb$_{1-x}$Sn$_x$Se (x=0.23-0.25) is very different from $Z_2$ topological insulators; in the latter materials $\lambda_{SS} \ll \delta$ and the IR data can be analyzed using a multilayer model with distinct surface and bulk layers[45, 47].

**Experimental uncertainty analysis**

The uncertainties of the bandgap $\Delta$ and Fermi velocity $v_F$ of the bulk states are estimated based on the variations of their values from fitting different LL transitions and a confidence interval of 95% in the least squares fit. The uncertainty of $E_F$ of the bulk states is estimated from the uncertainties in defining $E_{inter}$ in the $\sigma_1(\omega)$ spectrum or the magnetic field below which the transition $LL_{-0} \rightarrow LL_{+1}$ disappears. The uncertainty of the total Drude spectral weight $SW_{total}$ is discussed in Supplementary Discussion 5. In the analysis for data in magnetic field, the uncertainties of the spectral weight $SW_{SS}$ and scattering rate $1/\tau_{SS}$ of the CR mode of the SS represent the range of these parameters that can be used to reproduce the $R(\omega, B)$ data. The uncertainties of all other quantities discussed in the main text are calculated using standard formulas for propagation of uncertainty.

**Data availability**

The data that support the findings of this study are available from the corresponding author upon reasonable request.

**Acknowledgements**

Z.L. acknowledges support from 1000 young talents program of China and that of Sichuan province. J.L. and L.F. are supported by the DOE Office of Basic Energy Sciences, Division of Materials Sciences and Engineering under Award DESC0010526. F.C. acknowledges the support provided by the Ministry of Science and Technology (MOST) in Taiwan under Grant No. 104-2119-M-002 -028 -MY2. N.-L. W. acknowledges support from National Science Foundation of China (No. 11327806) and the National Key Research and Development Program of China (No.2016YFA0300902). A portion of this work was performed in N.-L. W.'s lab at Institute of Physics, Chinese Academy of Sciences. A portion of this work was performed at the National High Magnetic Field Laboratory, which is supported by National Science Foundation Cooperative Agreement No. DMR-1157490 and the State of Florida.

**Author contributions**

L.F. and Z.L. conceived the research and supervised the project together with F.C.; R.S. and F.C. contributed to the single-crystal growth; Z.L. and N.-L.W. supervised the experiments; Y.W. and Z.L. analyzed the data, interpreted the experimental results and wrote the paper; N.-L.W., L.F., G.L. and J.L. contributed to the data analysis. All authors discussed the results and commented on the paper.

**Competing financial interests**

The authors declare no competing financial interests.

insulators. Phys. Rev. B 89, 075138 (2014).

# Figures and Figure Legends

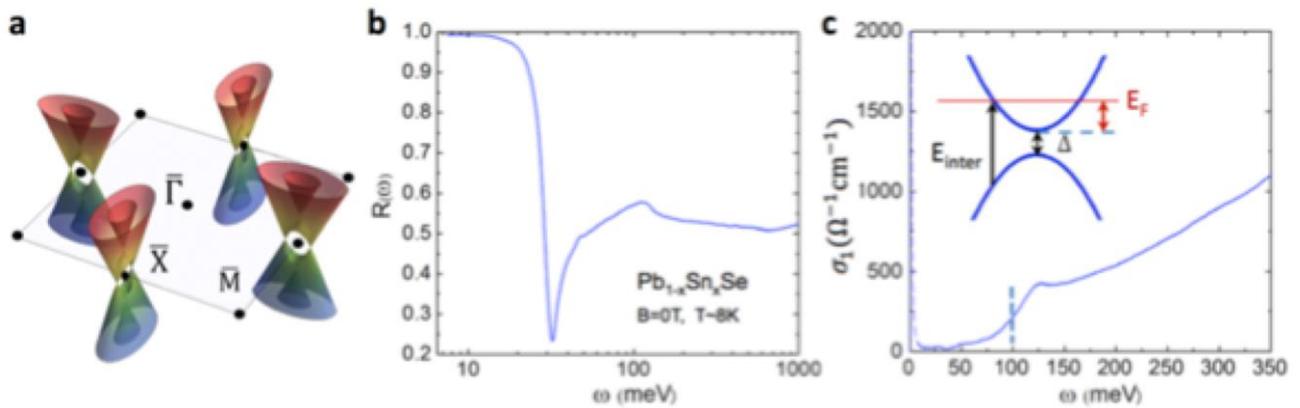

**Figure 1 | IR spectra of Pb$_{1-x}$Sn$_x$Se (x=0.23-0.25) in zero magnetic field. a**, Schematic band structure of the surface states and the surface Brillouin zone. **b**, IR reflectance spectrum $R(\omega)$ at T=8K. **c**, The real part of the optical conductivity $\sigma_1(\omega)$ at T=8K. The vertical dashed line around 100meV indicates E$_{inter}$. The inset shows a schematic of the band structure of the bulk states and the onset of interband transitions at E$_{inter}$.

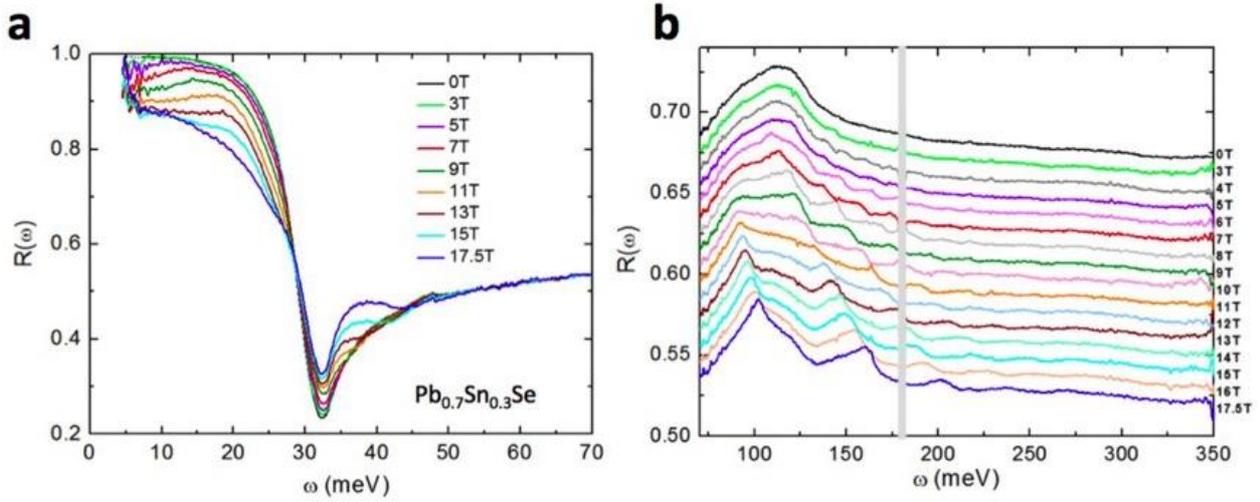

**Figure 2 | IR reflectance spectra in magnetic fields. a**, **b**, IR reflectance spectra $R(\omega, B)$ in several magnetic fields at T~4.5K. For clarity the spectra in **b** are displaced from one another by 0.01 with the spectrum at $B$=17.5T shown at its actual value. The gray area around 175 meV is the energy range in which no data can be obtained due to the IR absorption of the optical window in our setup.

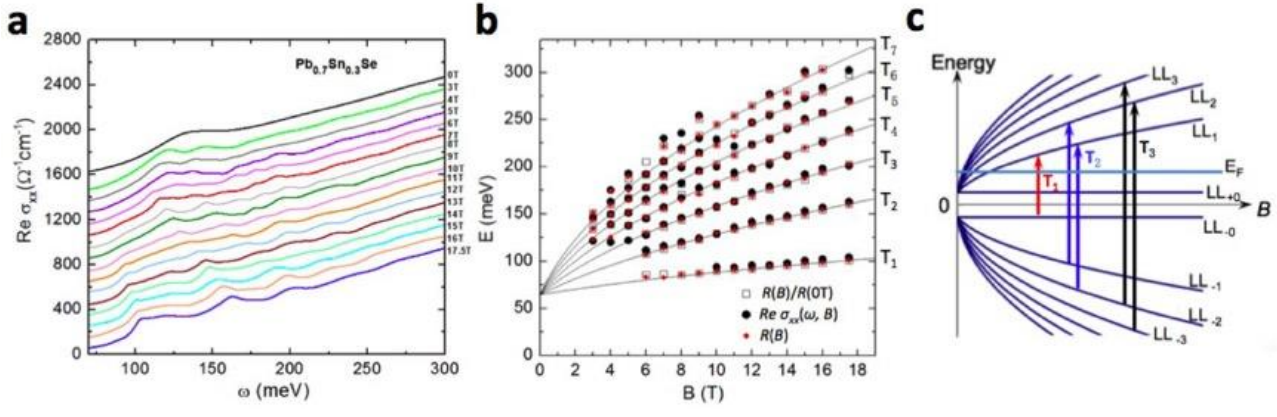

**Figure 3 | Landau level transitions of bulk states. a**, The real part of the optical conductivity $Re\ \sigma_{xx}(\omega)$ in several magnetic fields. For clarity the spectra are displaced from one another by 100 with the spectrum at $B$=17.5T shown at its actual value. **b**, The energies of the observed transitions as a function of magnetic field. Symbols: data. Solid lines: fits to the data using equation (1) discussed in the text with $\Delta$= 64.5 meV and $v_F = 0.4 \times 10^6$ m s$^{-1}$. The observed resonances can be assigned to LL transitions T$_n$, which is due to transitions $LL_{-(n-1)} \to LL_n$ and $LL_{-n} \to LL_{n-1}$ for n>1 and $LL_{-0} \to LL_1$ for n=1. **c**, Schematic of LLs of the bulk states. The arrows illustrate T$_1$-T$_3$ shown in **b**.

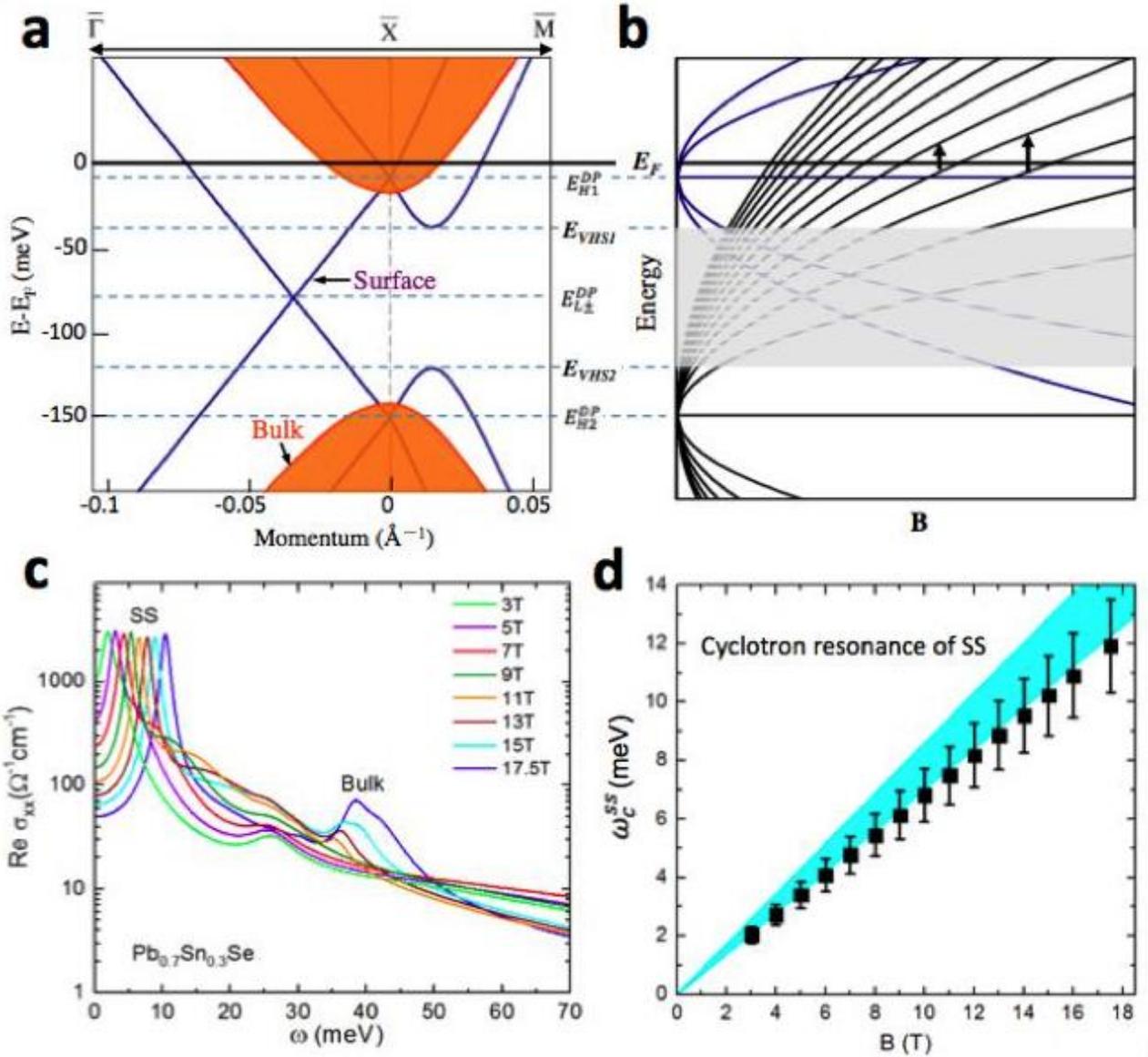

**Figure 4 | IR conductivity of surface states in magnetic fields. a**, A schematic of the SS band structure (dark blue) and that of the bulk (red) along two high-symmetry directions for one of the four Dirac cones inside the surface Brillouin zone. $E_{H1}^{DP}$, $E_{H2}^{DP}$ and $E_{L\pm}^{DP}$ are energies of the Dirac points associated with the two Dirac cones located at the $\bar{X}$ points and the Dirac nodes along the $\overline{\Gamma X}$ line, respectively. $E_{VHS1}$ and $E_{VHS2}$ are energies of the two Van Hove singularities in the band structure. **b**, A schematic of LLs of the SS in the same energy scale as that in **a**. While the LLs between $E_{VHS1}$ and $E_{VHS2}$ in the grey shaded area have nontrivial dispersions (ref. 32) (not shown here), the LLs near $E_F$ are well approximated by LLs of two independent Dirac cones at $E_{H1}^{DP}$ and

$E_{H2}^{DP}$ as illustrated by the dashed lines in the shaded area. The CR of the SS associated with the Dirac cone at $E_{H2}^{DP}$ is shown by the vertical arrows. **c**, The real part of the model optical conductivity $Re\ \sigma_{xx}(\omega)$ in several magnetic fields that are used to simulate the *R(ω, B)* spectra below 70 meV. The strong peak below 15 meV in $Re\ \sigma_{xx}(\omega)$ arises from CR of the SS, which represents the spatially averaged 3D optical conductivity of the SS within the IR penetration depth. **d**, The CR energy of SS $\omega_c^{SS}$ as a function of magnetic field. The error bars represent the range of $\omega_c^{SS}$ that can be used in the model $Re\ \sigma_{xx}(\omega)$ spectra to reproduce *R(ω, B)* as discussed in Supplementary Discussion 2. The shaded area indicates the range of $\omega_c^{SS}$ estimated from previous STM and ARPES experiments.

# Supplementary Information

# Observation of Ultrahigh Mobility Surface States in a Topological Crystalline Insulator by Infrared Spectroscopy

Ying Wang[1], Guoyu Luo[2], Junwei Liu[3], R. Sankar[4], Nan-Lin Wang[5], Fangcheng Chou[4], Liang Fu[3], Zhiqiang Li[2]*

[1]National High Magnetic Field Laboratory, Tallahassee, Florida 32310, USA.

[2]College of Physical Science and Technology, Sichuan University, Chengdu, Sichuan 610064, China.

[3]Department of Physics, Massachusetts Institute of Technology, Cambridge, Massachusetts 02139, USA.

[4]Center for Condensed Matter Sciences, National Taiwan University, Taipei 10617, Taiwan.

[5]International Center for Quantum Materials, School of Physics, Peking University, Beijing 100871, China.

*e-mail: zhiqiangli@scu.edu.cn

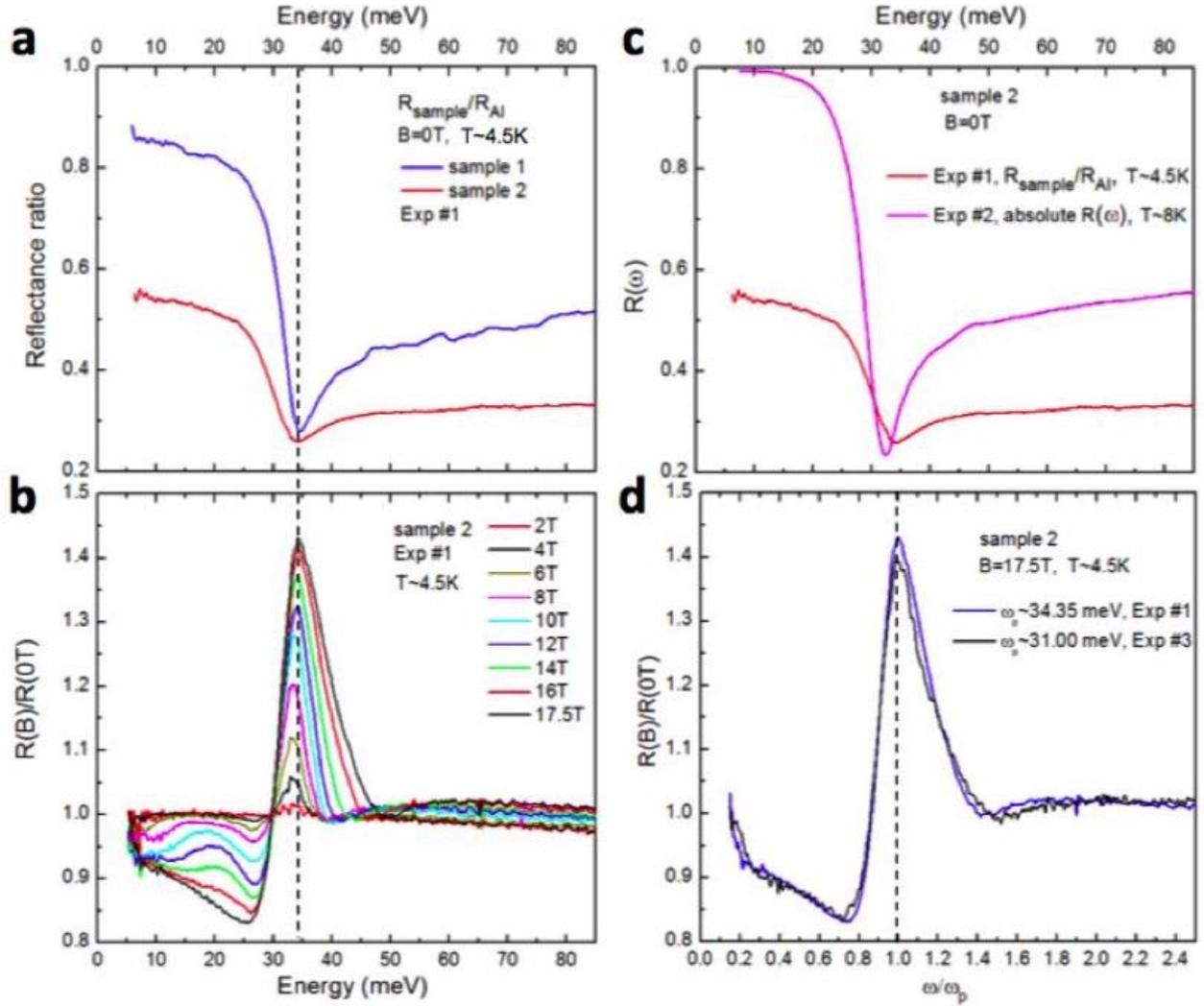

Figure S1: **a**, the reflectance ratio spectra $R_{sample}/R_{Al}$ for two $Pb_{1-x}Sn_xSe$ samples at zero field obtained in Exp #1. **b**, the magneto-reflectance ratio spectra $R(\omega, B)/R(\omega, B=0)$ for sample 2 obtained in Exp #1 (sample 1 shows similar results). The dashed line in **a** and **b** shows the screened plasma frequency $\widetilde{\omega}_P = \omega_P/\sqrt{\varepsilon_\infty}$. **c**, the reflectance ratio spectrum $R_{sample}/R_{Al}$ obtained in Exp #1 and absolute $R(\omega)$ spectrum obtained in Exp #2 for sample 2 in zero field. **d**, the magneto-reflectance ratio spectra $R(\omega, B)/R(\omega, B=0)$ at a representative field (B=17.5T) for sample 2 in Exp #1 and #3 displayed in a $\omega/\widetilde{\omega}_P$ plot.

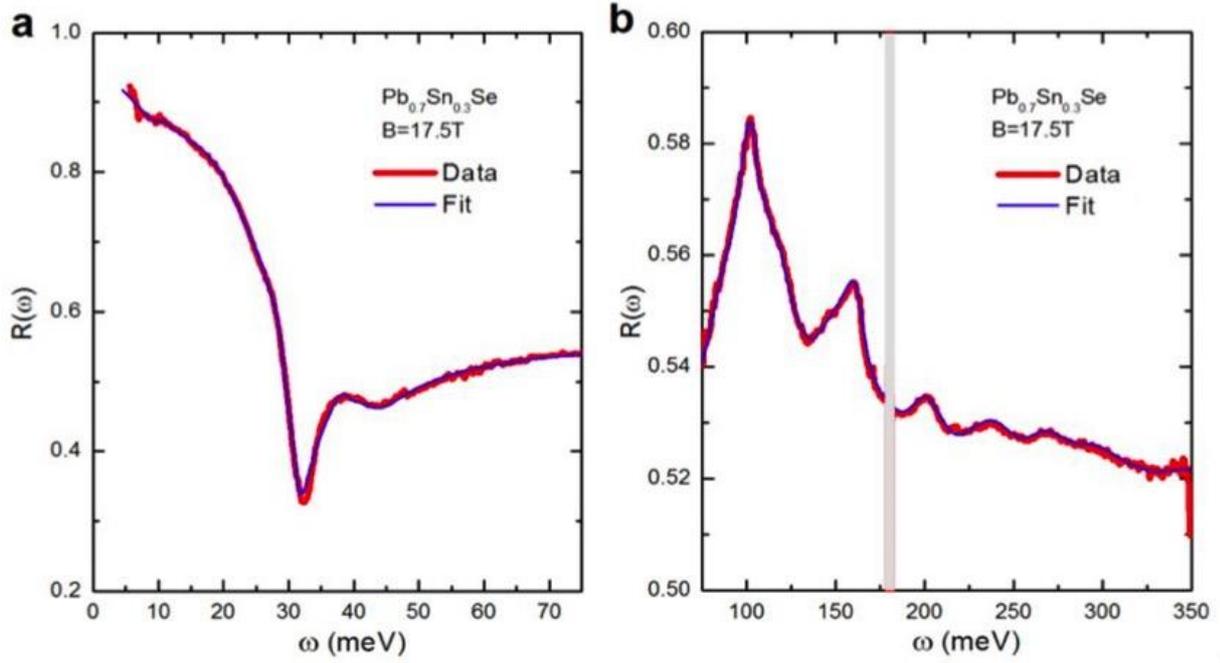

Figure S2: Experimental $R(\omega, B)$ spectra and model spectra at $B$=17.5 T. The spectra Fit #1 and #2 in panel **a** are obtained using model #1 and #2 shown in figures S3 and S4, respectively. Model #1 is the model presented in the main text. The fit spectra in panel **b** are obtained using model #1. The vertical axis is shown in different scales in **a** and **b** to highlight small features in $R(\omega, B)$. The gray area around 175 meV is the energy range in which no data can be obtained due to the IR absorption of the optical window in our setup.

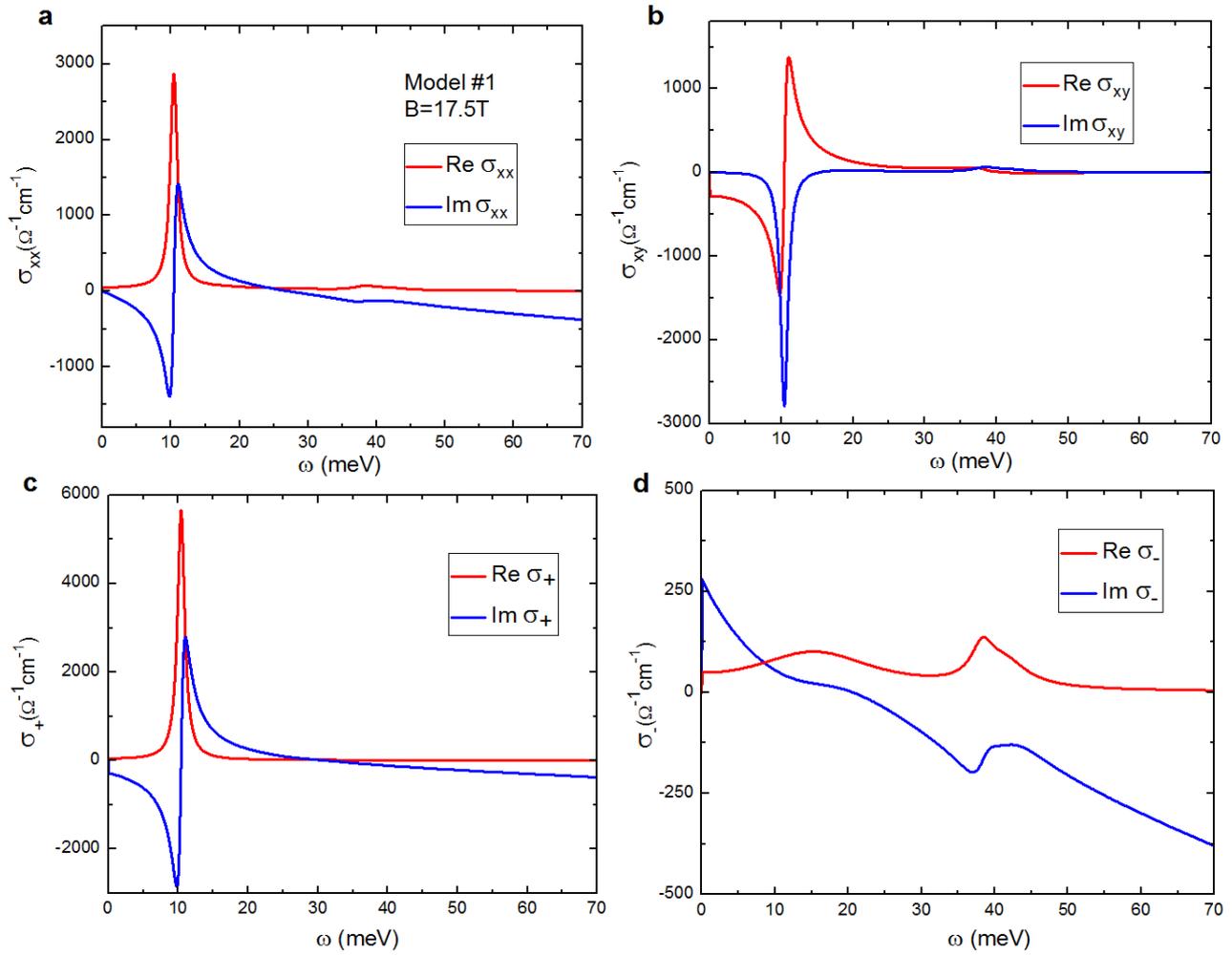

Figure S3: The real and imaginary parts of $\sigma_{xx}(\omega)$, $\sigma_{xy}(\omega)$, $\sigma_+(\omega)$ and $\sigma_-(\omega)$ for model #1 with all parameters summarized in table S1, which is the model presented in the main text.

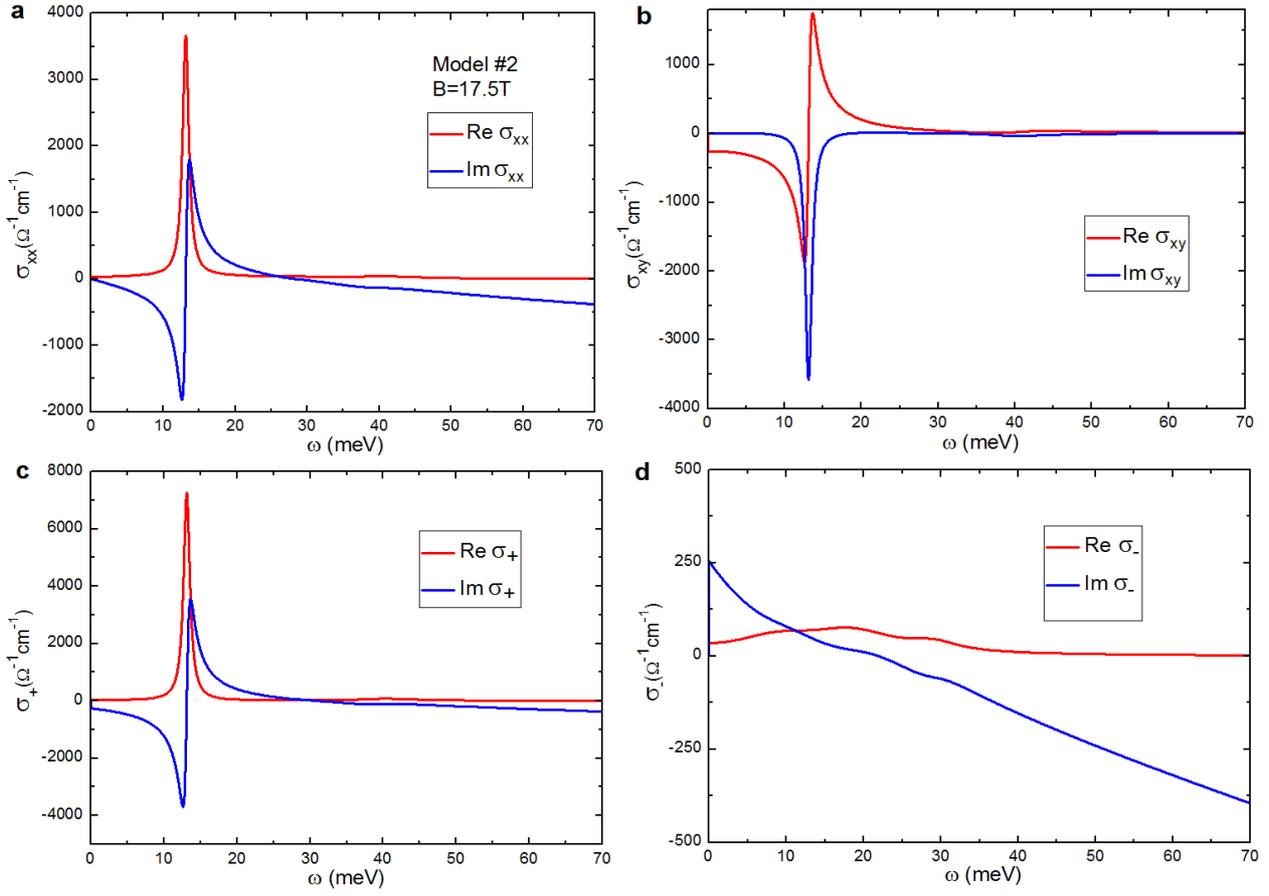

Figure S4: The real and imaginary parts of $\sigma_{xx}(\omega)$, $\sigma_{xy}(\omega)$, $\sigma_{+}(\omega)$ and $\sigma_{-}(\omega)$ for model #2 with all parameters summarized in table S2.

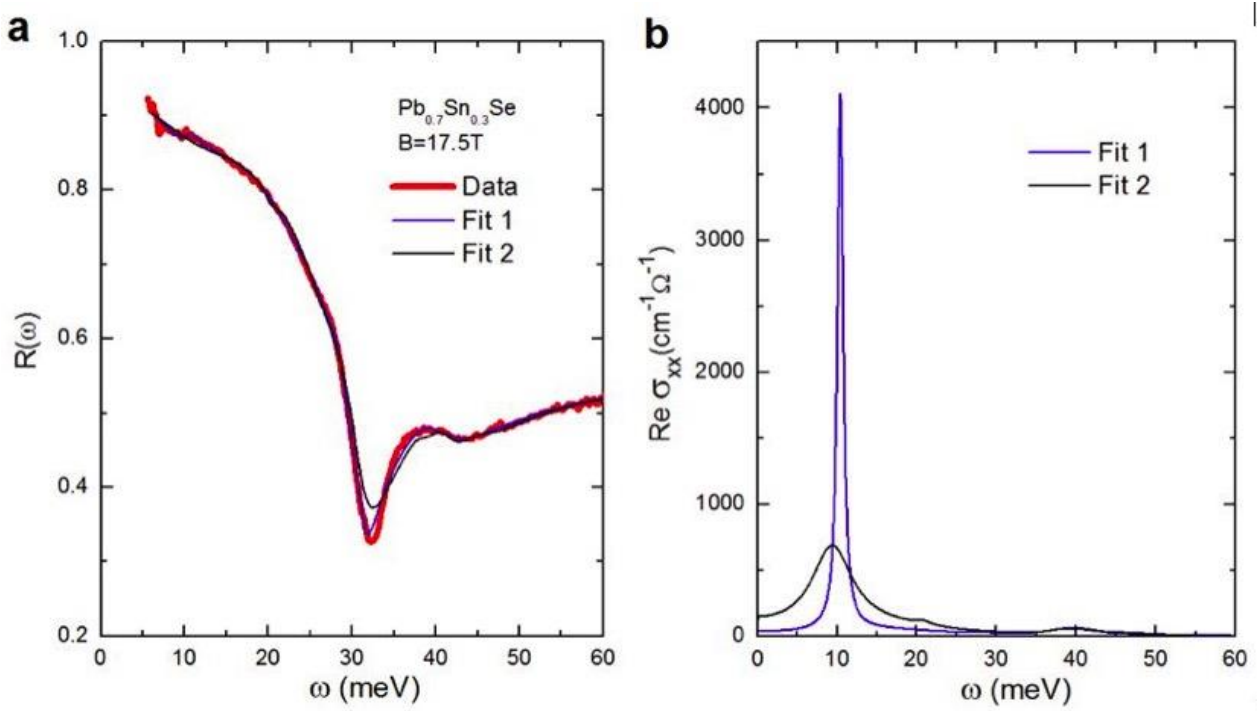

Figure S5: **a**, Experimental $R(\omega, B)$ spectra and two model spectra at $B$=17.5 T. **b**, The $Re\ \sigma_{xx}(\omega, B)$ spectra in two models used to generate the model spectra in **a**. Blue curves: $1/\tau_{SS}$=0.6 meV. Black curves: $1/\tau_{SS}$=3meV.

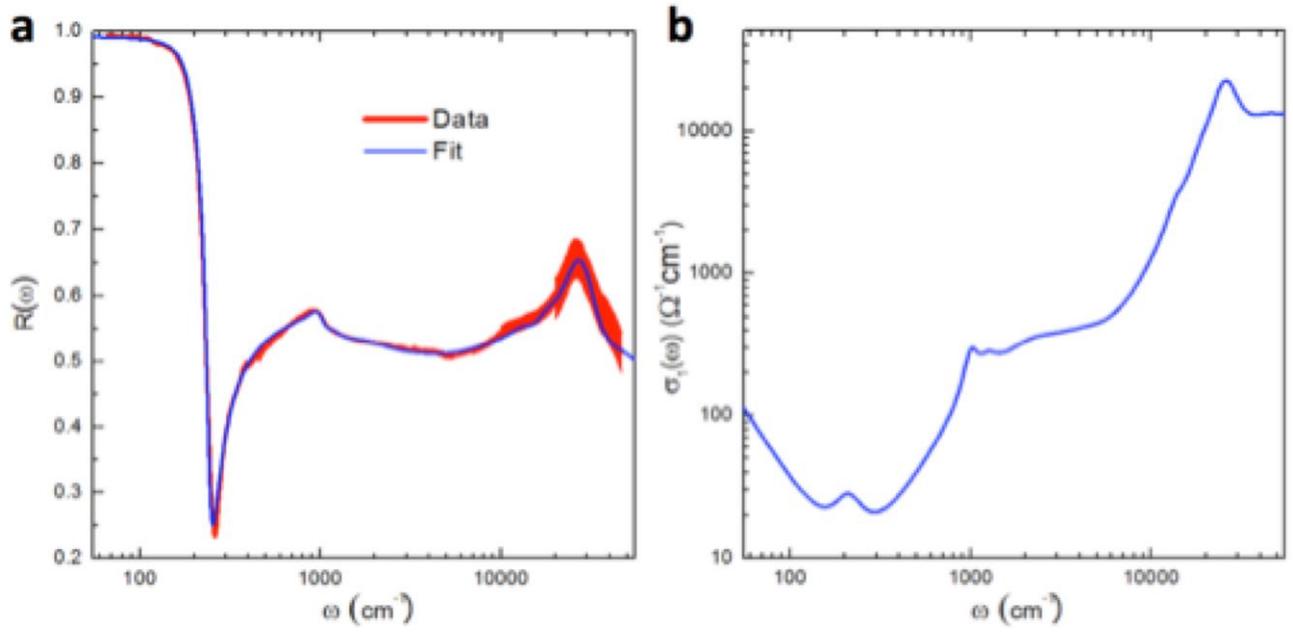

Figure S6: **a**, Experimental $R(\omega)$ spectrum in zero field and a typical fit using the Drude-Lorentz model. The thickness of the experimental spectrum represents the estimated uncertainty. **b**, the corresponding $\sigma_1(\omega)$ spectrum. The parameters for all oscillators in the model are summarized in Table S3.

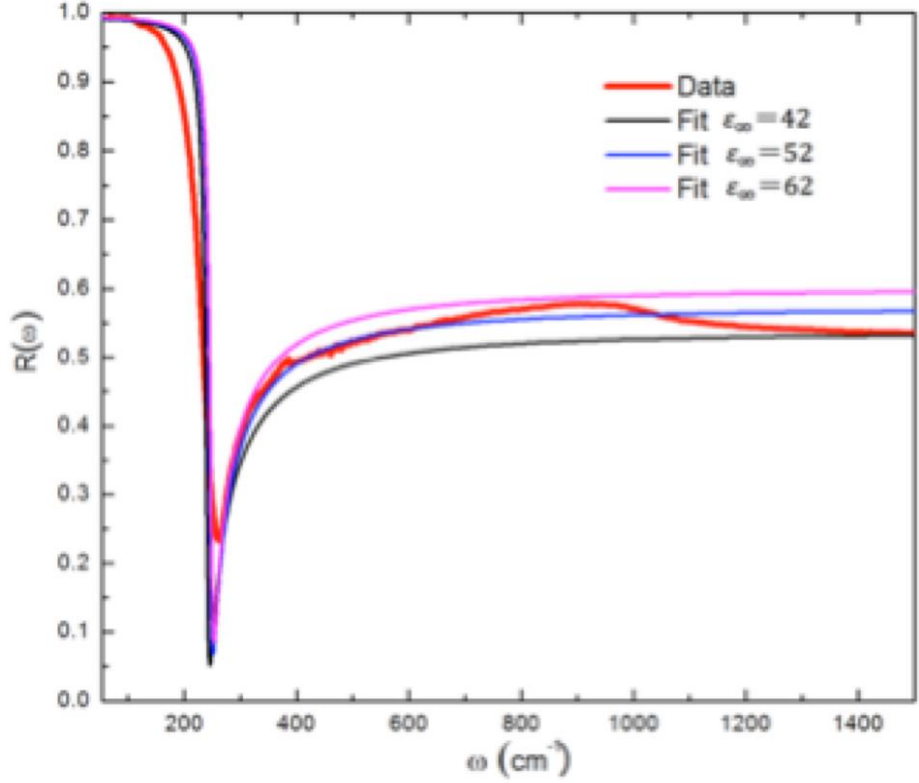

Figure S7: The experimental $R(\omega)$ spectrum in zero field and model $R(\omega)$ spectra calculated from the Drude model $\sigma(\omega) = \frac{\omega}{4\pi i}(\varepsilon_\infty - 1 - \frac{\omega_p^2}{\omega^2 + i\gamma\omega})$, where $\varepsilon_\infty$ represents all electronic contributions to the dielectric constant other than the Drude conductivity. The model spectra with different values of $\varepsilon_\infty$ are generated using $\gamma = 7$ cm$^{-1}$ and $\omega_P = \widetilde{\omega}_P\sqrt{\varepsilon_\infty}$ with $\widetilde{\omega}_P \sim 260$ cm$^{-1}$. The $R(\omega)$ spectrum above $\widetilde{\omega}_P$ increases with increasing value of $\varepsilon_\infty$.

Table S1: Parameters for optical conductivity model spectra for B=17.5T presented in the main text and Figure S3 (model #1). These parameters are used in Eq. (3) of the main text. $\varepsilon_\infty^B = 49.24$, and $\omega_{o,n} = 0$ for all oscillators.

| Energy $\omega_{c,n}$ (cm$^{-1}$) | Plasma frequency $\omega_{p,n}$ (cm$^{-1}$) | Linewidth $\gamma_n$ (cm$^{-1}$) |
|---|---|---|
| 124 | 689 | 82 |
| 309 | 276 | 15 |
| 334 | 325 | 30 |
| -84 | 1301 | 5 |
| -218 | 1117 | 19 |
| -244 | 81 | 12 |

Table S2: Parameters for optical conductivity model spectra for B=17.5T presented in Figure S4 (model #2). These parameters are used in Eq. (3) of the main text for magneto-Drude-Lorentz model #2, which is another representative model. $\varepsilon_\infty^B = 49.73$, and $\omega_{o,n} = 0$ for all oscillators.

| Energy $\omega_{c,n}$ (cm$^{-1}$) | Plasma frequency $\omega_{p,n}$ (cm$^{-1}$) | Linewidth $\gamma_n$ (cm$^{-1}$) |
|---|---|---|
| 70 | 338 | 50 |
| 148 | 459 | 59 |
| 235 | 200 | 30 |
| -106 | 1351 | 4.2 |
| -230 | 151 | 25 |
| -325 | 350 | 30 |
| -370 | 226 | 55 |

**Table S3: Parameters for Drude-Lorentz model spectra for B=0T presented in Figure S6.** These parameters are used in Eq. (3) of the main text with $\omega_{c,n} = 0$ for all oscillators, which is the zero-field Drude-Lorentz model. $\varepsilon_\infty^B = 2.733$.

| Energy $\omega_{0,n}$ (cm⁻¹) | Plasma frequency $\omega_{p,n}$ (cm⁻¹) | Linewidth $\gamma_n$ (cm⁻¹) |
|---|---|---|
| 0 | 1730.98 | 7 |
| 209.78 | 276.55 | 93.41 |
| 1005.90 | 1325.32 | 201.24 |
| 1241.20 | 1326.11 | 382.13 |
| 2313.25 | 7785.34 | 3572.83 |
| 4310.72 | 4058.41 | 3426.14 |
| 13800 | 14239.48 | 4050 |
| 18960.71 | 24857.39 | 5535.32 |
| 25645.54 | 104447.11 | 10000.87 |
| 44100.83 | 139379.84 | 39529.24 |
| 71803.38 | 164841.51 | 57286.62 |

## 1. *R(ω, B)* spectra of Pb$_{1-x}$Sn$_x$Se and discussion on Se vacancy defects

The magneto-reflectance ratio *R(ω, B)/R(ω, B=0)* and the zero-field reflectance *R(ω)* for two Pb$_{1-x}$Sn$_x$Se (x=0.23-0.25) samples were measured in 3 different experiments (Exp #1-3 in sequence) with several months between one another. All the data in the main text are from sample 2. Figure S1a shows the reflectance ratio data R$_{sample}$/R$_{Al}$ at zero field obtained in Exp #1, which is the reflectance of sample divided by that of a reference aluminum mirror. Because of the lack of in situ gold coating in this experiment, the R$_{sample}$/R$_{Al}$ spectra are not *absolute R(ω)*. Nevertheless, the plasma minimum (dip in R$_{sample}$/R$_{Al}$) at the screened plasma frequency $\widetilde{\omega}_P = \omega_P/\sqrt{\varepsilon_\infty}$ shown in figure S1a should be fairly accurate owing to the frequency independent *R(ω)* of aluminum in this range. Here, the bare plasma frequency $\omega_P$ is given by $\omega_P^2 = 4\pi e^2 n/m$, $\varepsilon_\infty$ represents all high-energy contributions to the dielectric constant other than the free carrier contribution. The data in figures S1a and S1b show that the peak in the *R(ω, B)/R(ω, B=0)* spectrum is exactly at the plasma minimum frequency $\widetilde{\omega}_P$ at zero field measured in the same experiment.

In order to obtain the *absolute R(ω)* spectrum at zero field, we performed Exp #2 with in situ gold coating technique, after which *R(ω, B)/R(ω, B=0)* were measured again in Exp #3. We find that $\widetilde{\omega}_P$ changes slightly from ~34.35 meV to ~32.49 meV (figure S1c), and then to ~31 meV in Exp #1-3. This observation probably arises from self-doping due to Se vacancies caused by the so-called "Se loss phenomenon" in selenides, namely, thermal energy assisted surface Se atom escape, which is very common in selenides such as TiSe$_2$, SnSe and Bi$_2$Se$_3$ as discussed in ref. [1] and references therein. The sample is cooled down to ~4.5K or ~10K then warmed up to 300K for multiple times during different measurements, between which it is briefly exposed to air. Every thermal cycling could introduce Se loss, especially for the Se atoms at the surface with the weakest chemical bonding. The Se loss effect could lead to changes in the carrier density n and therefore the change of $\tilde{\omega}_P$ with time, since $\tilde{\omega}_P \propto \sqrt{n}$. Importantly, we find that the frequencies of the features in $R(\omega, B)/R(\omega, B=0)$ below 60 meV change proportionally with $\tilde{\omega}_P$. As shown in figure S1d, $R(\omega, B)/R(\omega, B=0)$ measured in Exp #1 and #3 remain the same within experimental uncertainty (1%-1.5%) if they are plotted in a $\omega/\tilde{\omega}_P$ plot. The scaling results in figure S1d demonstrate the *robustness* of the observed features in $R(\omega, B)/R(\omega, B=0)$ despite the small change of $\tilde{\omega}_P$ with time, which is an extrinsic effect. Also, the observed main features in $R(\omega, B)/R(\omega, B=0)$ above 60 meV change little from Exp #1 to Exp #3, which we believe is due to the fact that these features arise from interband Landau level transitions and are therefore relatively insensitive to the change of carrier density.

There are two methods to account for the small change of $\tilde{\omega}_P$ with time in our data:

(1) One can use the absolute $R(\omega, B=0)$ spectrum obtained in Exp #2 (figure S1c), and use the $R(\omega, B)/R(\omega, B=0)$ spectra in Exp #1 (figure S1b) after scaling the frequency by the ratio of the $\tilde{\omega}_P$ in Exp #2 and #1 below 60 meV. The validity of this approach is justified by the scaling results shown in figure S1d, especially because Exp #2 is performed between Exp #1 and #3.

(2) The $R(\omega, B)/R(\omega, B=0)$ in Exp #1 will be used. Then the corresponding $R(\omega, B=0)$ spectrum can be obtained from that in Exp #2 (figure S1c) after scaling the frequency by the ratio of the $\tilde{\omega}_P$ in Exp #1 and #2 below 60 meV. This is justified because the $R(\omega, B=0)$ spectrum obtained in Exp #2 can be satisfactorily described by the Drude model, and the $R(\omega)$ spectra in the Drude model with different $\tilde{\omega}_P$ will be identical if they are all displayed a $\omega/\tilde{\omega}_P$ plot.

In both methods, the $R(\omega, B)$ spectra are obtained by multiplying the $R(\omega, B)/R(\omega, B=0)$ spectra by $R(\omega, B=0)$. In the main text, data obtained using the first method is presented and discussed. The data obtained from the second method yield the same conclusions discussed in the main text.

It has been a longstanding challenge in the research of selenides to control of the sample quality. Even for samples with the same nominal Pb/Sn ratio, the actual carrier density is sample dependent, which depends strongly on the self-doping effect of Se vacancy defects in each crystal. In the Bridgman crystal growth, the details can be slightly different for every batch, such as vacuum control and the specific solidification temperature at the solid-liquid interface. Moreover, the vacancy defect has higher density closer to the surface. All of these factors fluctuate in each growth, which leads to sample-dependent defect density and therefore carrier density. The samples used in our IR study are from a batch with low carrier density, but the carrier density is batch dependent even for the same nominal Pb/Sn ratio.

## 2. Details on $\sigma_{xx}(\omega)$ and $\sigma_{xy}(\omega)$ in our analysis of $R(\omega, B)$ and CR mode of the surface states

As discussed in the main text, the reflectance data $R(\omega, B)$ were analyzed using the magneto-Drude-Lorentz model. This model ensures that the real and imaginary parts of optical conductivity are constrained by Kramers–Kronig relations, therefore it is commonly used to describe Landau level transitions including cyclotron resonance (CR). This model is among the most used methods of parametrization of optical response functions including optical conductivity. The physical meaning of each individual oscillator in the model may not be always clear, but the meaningful result is the *overall* optical conductivity obtained from *all* oscillators, as discussed in details in reference 28 of the main text and references therein. In our analysis, we use one oscillator at $\omega_c^{SS} = eB/m_{ss}$ based on theoretical study of Landau levels of the SS in TCIs (reference 32 of the main text) and several much weaker oscillators to simulate the $R(\omega, B)$ spectra below 60 meV, which represent a parametrization of the overall optical conductivity as explained above. Figure S2 displays the experimental $R(\omega, B)$ spectra together with two representative model spectra at $B$=17.5

T. The real and imaginary parts of $\sigma_{xx}(\omega)$, $\sigma_{xy}(\omega)$, $\sigma_+(\omega)$ and $\sigma_-(\omega)$ in representative models are shown in Fig. S3 and S4, where model #1 is corresponding to the model for B=17.5T in Fig 4c of the manuscript. The parameters for all oscillators in each model are summarized in table S1 and S2.

Depending on the width and spectral weight of the oscillators used in our analysis and their distributions in $\sigma_+(\omega)$ and $\sigma_-(\omega)$, we find that a range of $\omega_c^{ss}$ values with different possible $\sigma_{xx}(\omega)$ spectra can reproduce the spectral feature in $R(\omega, B)$ below 25 meV and its evolution with $B$ field. Two representative models are shown in Fig. S3 and S4. Specifically, simulated $\sigma_{xx}(\omega)$ spectra using $m_{ss}$ values in the range of 0.15-0.19 $m_e$ for the CR mode yield model $R(\omega, B)$ spectra that can fit the experimental data with the same quality as those shown in figure S2a within experimental uncertainties. As discussed in the main text, scanning tunneling microscopy (STM) and angle-resolved photoemission spectroscopy (ARPES) experiments [2-4] suggest that the effective mass for the SS associated with the Dirac cone at $E_{H2}^{DP}$ is $m_{ss} = E_F^{SS}/(\bar{v}_F^{SS})^2$ =0.15±0.015 $m_e$. Therefore the effective mass values $m_{ss}$ estimated from our IR experiments are consistent with those inferred from STM and ARPES measurements within 15% (Fig. 4d of the main text). This excellent agreement strongly supports our identification of the CR mode of SS discussed in the main text. The small deviation between our results and STM and ARPES measurements may arise from the difference in Fermi energy in different samples. Moreover, the spectral feature in $R(\omega, B)$ data below 25 meV are well below the energy range of ALL allowed LL transitions from the bulk states, so it can only be assigned to the SS.

The lineshape of $R(\omega, B)$ is determined by real and imaginary parts of $\sigma_{xx}(\omega)$ and $\sigma_{xy}(\omega)$ based on Eq.(4) in the main text. Although there is a sharp peak due to SS in $Re\ \sigma_{xx}(\omega)$, the contributions from ALL real and imaginary parts of $\sigma_{xx}(\omega)$ and $\sigma_{xy}(\omega)$ in Fig. S3 and S4 lead to a relatively smooth lineshape for $R(\omega, B)$ in this spectral range, which is a consequence of the spectral properties (lineshapes) of optical response functions typically described by magneto-Drude-Lorentz oscillators.

**3. Comparing $R(\omega, B)$ data with the conventional magnetoplasma effect in semiconductors**

Conventional doped semiconductors exhibit the so-called magnetoplasma effect in magnetic field [5, 6]: the plasma minimum (edge) in the $R(\omega)$ spectra in zero field splits into two minimums (edges) in $R(\omega, B)$ in magnetic field, with the high (low) energy edge moving to higher (lower) energies with increasing field. This effect is observed in the regime of $\omega_c \ll \widetilde{\omega}_P$, where $\widetilde{\omega}_P$ is the frequency of the plasma minimum in the $R(\omega)$. Such a behavior in $R(\omega, B)$ spectrum arises from a *single* CR mode at $\omega_c$ in the optical conductivity $\sigma_{xx}(\omega, B)$ that is well separated in energy from other resonances such as interband LL transitions [5, 6]. In our measurements of $Pb_{1-x}Sn_xSe$, the observed features in $R(\omega, B)$ below 50 meV are entirely different from the magnetoplasma effect described above. Our observation results from the overall contributions of *two* resonances in this energy range: the CR mode of the SS and the $LL_{+0} \to LL_{+1}$ transition from the bulk states (Fig. 4c of the main text). We stress that the $LL_{+0} \to LL_{+1}$ transition around 30-40 meV is the LL transition with the lowest energy from the bulk states, which can not produce the dramatic changes below ~25 meV in $R(\omega, B)$ with increasing magnetic field. The latter observation suggests the existence of a low energy resonance below the $LL_{+0} \to LL_{+1}$ transition of the bulk, which we demonstrate to be the CR mode of the SS.

**4. Scattering rate and mobility of the surface states**

As illustrated in figure S5, the width of the dip feature around 32 meV in $R(\omega, B)$ is directly related to the width of the CR mode of the SS in $Re\ \sigma_{xx}(\omega, B)$ spectra. Therefore, the very narrow dip feature around 32 meV in the $R(\omega, B)$ data suggests that the scattering rate for the SS $1/\tau_{SS}$ is very low. We find that $Re\ \sigma_{xx}(\omega, B)$ spectra with $1/\tau_{SS} \sim 1.2 \pm 0.6$ meV for the surface CR mode can reproduce the $R(\omega, B)$ data. As shown in Fig. S5, larger values of $1/\tau_{SS}$ are inconsistent with the narrow dip feature in $R(\omega, B)$. This estimation of $1/\tau_{SS}$ allows us to estimate the mobility of the SS as discussed in the main text.

It is instructive to compare the surface mobility in TCIs to the mobility of graphene [7-9] since both materials feature massless Dirac fermions. In a simple Drude model, the mobility of carriers in graphene is given by [8]:

$$\mu = ev_F\tau/(\hbar k_F) = e\tau/m, \qquad (S1)$$

where $m = E_F/v_F^2$. In a previous study [9] of graphene samples with carrier density $n \sim 4.7 \times 10^{12}$ cm$^{-2}$ (corresponding to $m \sim 0.04$ m$_e$), the scattering rate is found to be $1/\tau \sim 30$ meV based on IR study of Landau level transitions, which yields an IR mobility of $\mu_{IR} \sim 5{,}700$ cm$^2$ V$^{-1}$ s$^{-1}$ based on equation (S1). The estimated $\mu_{IR}$ is in reasonable agreement with the reported DC mobility $\mu_{DC} \sim 4{,}000$ cm$^2$ V$^{-1}$ s$^{-1}$ [9]. The lower value of $\mu_{DC}$ compared to $\mu_{IR}$ might arise from disorder effects as discussed in [7]. The good agreement between $\mu_{DC}$ and $\mu_{IR}$ in graphene [9] supports our estimation of the surface mobility $\mu_{SS}$ in TCIs using an equation similar to equation (S1). Moreover, a scattering rate of $1/\tau \sim 2$ meV was observed in IR measurements of graphene/BN with mobility $\mu \sim 50{,}000$ cm$^2$ V$^{-1}$ s$^{-1}$ [7]. Therefore, the scattering rate and mobility from previous studies of graphene [7,9] provide further support for our estimation of $\mu_{SS} \sim 40{,}000$ cm$^2$ V$^{-1}$ s$^{-1}$ in TCIs.

**5. Analysis of total Drude spectral weight in zero field**

We use the Drude-Lorentz model to fit the $R(\omega)$ data in zero field. One typical fit to $R(\omega)$ and the corresponding $\sigma_1(\omega)$ are shown in Fig. S6 with the parameters for all oscillators in the model summarized in Table S3. The optical conductivity $\sigma(\omega)=\sigma_1(\omega)+i\sigma_2(\omega)$ from Drude-Lorentz fit is consistent with that evaluated from Kramers-Kronig (KK) transformation of $R(\omega)$. The total Drude spectral weight ($SW_{total}$) can be directly determined by the bare plasma frequency $\omega_P$ of the Drude oscillator from $SW_{total} = \frac{\pi\Omega^{-1}}{120}\omega_P^2$, where $\omega_P$ is in cm$^{-1}$, $\Omega$ is Ohm. The total (observed) plasma frequency $\omega_P$ for the Drude mode can be obtained from *fitting the entire R(ω) spectrum* shown in Fig. S6a, because it is related to the screened plasma frequency $\widetilde{\omega}_P$ by:

$$\omega_P = \widetilde{\omega}_P \sqrt{\varepsilon_\infty} \qquad (S2)$$

where $\widetilde{\omega}_P \sim 260$ cm$^{-1}$ corresponds to the plasma minimum in $R(\omega)$ and $\varepsilon_\infty$ is determined by all Lorentzian oscillators obtained from fitting the overall R(ω) spectrum *above* $\widetilde{\omega}_P$. $\varepsilon_\infty$ represents all electronic contributions to the dielectric constant other than the Drude conductivity, which is given by (see for example, [10]):

$$\varepsilon_\infty \equiv \varepsilon_1(\omega)|_{\omega \to 0} = 1 + \frac{120}{\pi} P \int_{\omega_0}^{\infty} \frac{\sigma_1(\omega')}{\omega'^2 - \omega^2} d\omega' \Big|_{\omega \to 0} \qquad (S3)$$

where P denotes the Cauchy principal value, frequencies are in cm$^{-1}$, $\sigma_1$ is in $\Omega^{-1}$cm$^{-1}$, and $\omega_0$ is a frequency separating the Drude mode and the interband transitions ($\omega_0 \sim 200$ cm$^{-1}$). The high energy contribution of $\sigma_1(\omega)$ to $\varepsilon_\infty$ is negligibly small because of the denominator of the integrand in Eq. (S3). Taking into account the uncertainties of $\sigma_1(\omega)$ from both Drude-Lorentz fit and KK transformation of $R(\omega)$ as well as those for $\omega_0$, we estimate $\varepsilon_\infty \sim 45 \pm 9$ using Eq. (S3).

In fact, the lineshape and absolute value of the $R(\omega)$ spectrum *above* $\widetilde{\omega}_P$ directly reflects the value of $\varepsilon_\infty$. Because the $R(\omega)$ spectrum shows a typical metallic behavior near and below $\widetilde{\omega}_P$, we can use a simple Drude model to illustrate the dependence of $R(\omega)$ on $\varepsilon_\infty$. Fig. S7 displays model $R(\omega)$ spectra calculated from the Drude model $\sigma(\omega) = \frac{\omega}{4\pi i}(\varepsilon_\infty - 1 - \frac{\omega_p^2}{\omega^2 + i\gamma\omega})$ with different values of $\varepsilon_\infty$, which shows that the $R(\omega)$ spectrum above $\widetilde{\omega}_P$ increases with increasing value of $\varepsilon_\infty$. Although $R(\omega)$ spectra from the Drude model in Fig. S7 only intends to qualitatively illustrate the effect of $\varepsilon_\infty$, it shows that the absolute value of the experimental $R(\omega)$ spectrum *above* $\widetilde{\omega}_P$ can be reproduced by $\varepsilon_\infty \sim 52 \pm 10$, which is in good agreement (within 15%) with the result from our full analysis ($\varepsilon_\infty \sim 45 \pm 9$) using Drude-Lorentz model. Therefore, our analysis demonstrates that $\omega_P = \widetilde{\omega}_P\sqrt{\varepsilon_\infty} \sim (1744 \pm 174)$ cm$^{-1}$ and $SW_{total} \approx (7.9 \pm 1.6) \times 10^4$ $\Omega^{-1}$cm$^{-2}$. This method of evaluating $SW_{total}$ doesn't rely on the integral $\int_0^{\omega_0} \sigma_1(\omega)d\omega$, so it is not limited by the lack of information on $\sigma_1(\omega)$ for the Drude mode below ~7meV.